\documentstyle[prb,aps,preprint]{revtex}
\begin{document}

\title{Ac susceptibility behaviour of Ce$_2$Pd$_{1-x}$Co$_{x}$Si$_3$}

\author{E.V. Sampathkumaran}

\address{Tata Institute of Fundamental Research, Homi Bhabha Road, Colaba, 
Mumbai - 400 005, INDIA.}

\maketitle

\begin{abstract}

We report the results of ac susceptibility ($\chi$)  measurements
for various compositions of the series,
Ce$_2$Pd$_{1-x}$Co$_x$Si$_3$, in the polycrystalline form, which has been previously reported to 
undergo a transformation from magnetic ordering to non-magnetic Kondo 
lattice behavior as x is varied from 0 to 1. A finding of emphasis here is that  
the features in ac $\chi$ data for
the compositions, x= 0.6 and 0.7, in the close vicinity of 
quantum critical point (QCP), are typical of 
long range magnetic ordering systems, whereas the compositions 
far away from QCP at the Pd-end including Ce$_2$PdSi$_3$
exhibit spin-glass characteristics. This trend is  
different from hitherto known behavior among Kondo alloys.

\end{abstract}


\vskip1cm
\newpage
\maketitle

It has been generally assumed in the literature 
that there may be an intimate relationship
between disorder, SG ordering and NFL behavior at QCP.  
In this article, we present ac magnetic susceptibility ($\chi$$_{ac}$)
behavior of a pseudoternary 
series, Ce$_2$Pd$_{1-x}$Co$_x$Si$_3$ (Ref. 2), crystallizing in 
a AlB$_2$-derived hexagonal structure. The results suggest that
the SG
behavior as revealed by the features in  the ac $\chi$ data  is actually seen only if one moves 
away from QCP in this series in contrast to hitherto known trends among
Kondo lattices.\cite{1} 

For previous work on these alloys, the readers may see Refs 1-5.
The results on the pseudoternary alloys 
revealed that there is a transformation from magnetic ordering 
for x= 0.0 to non-magnetic heavy fermion behavior for x= 1.0. 
The magnetic ordering temperature (T$_o$) falls\cite{7} 
in the region 2 to 4 K (Ref. 2) 
for x $<$ 0.7 reducing to a value  close to 2.3 K for x= 0.7;  
for x= 0.8, T$_o$ (if present) apparently dips to a value 
well below  2 K. Given\cite{2,7} that Ce$_2$CoSi$_3$ is non-magnetic, 
it is clear that QCP lies at a composition close to 0.8. In other words, gradual Co 
substitution for Pd results in a shift towards QCP in this alloy series.

The samples, Ce$_2$Pd$_{1-x}$Co$_x$Si$_3$ (x= 0.0, 0.2, 0.4, 0.6, 
0.7), employed here are the same as those in Ref. 1. The 
temperature (T) dependence (1.8 - 20 K) of real and imaginery 
parts of ac susceptibility ($\chi$$\prime$ and $\chi$$\prime\prime$ repectively) 
were obtained by a commercial superconducting
quantum interference device (Quantum Design) (ac driving field 1 Oe) at
several frequencies (1, 10, 100 and 1000 Hz). 

The results of our measurements are shown in Figs. 1 and 2. It is 
obvious from Fig. 1 that ac $\chi$$\prime$ tends to peak for all the 
compositions and the peak positions (about 2.7, 3.3, 3.3, 3.1 and 2.3 
for x= 0.0, 0.2, 0.4, 0.6 and 0.7 respectively) mark the appearance of 
magnetic ordering in all these alloys.\cite{6} For x= 0.7, the onset of magnetic
ordering at 2.3 K is manifested as a tendency  
of $\chi$ to flatten both 
in ac $\chi$$\prime$ (Fig. 1) and dc $\chi$ (Ref. 1). In order to understand whether
the magnetic transitions are of a long range  or of a SG character, it
is important to look at the frequency dependence of ac $\chi$$\prime$ as well as
the behavior of $\chi$$\prime$$\prime$.  It is now well established that 
long range magnetic ordering (LRMO) systems do not exhibit any frequency dependence of ac $\chi$
at T$_o$, unlike SG systems which show a marginal shift of the curves (as indicated
by a reduction of the $\chi$$\prime$ 
values below the peak temperature) towards high
temperature 
with increasing frequency with a well-defined cusp in the 
$\chi$$\prime$-T plot at T$_o$; in addition, $\chi$$\prime$$\prime$ 
of SG systems may exhibit a sharp
prominent upturn at T$_o$ as T is lowered, a feature that is absent in 
LRMO systems. With these criteria in mind, one can
make  relevant conclusions looking at Figs. 1 and 2. Though we have 
collected the data at four frequencies, for the sake clarity, we show the plots
in Fig. 1  at two frequencies only for x= 0 and 0.2. For x= 0.6 and 0.7, the plots
at the four frequencies overlap with each other even in the magnetically 
ordered state. Thus, these two 
alloys are definitely not spin-glasses; as a firm support, 
there is no upturn in $\chi$$\prime$$\prime$ at T$_o$; therefore, these
two compositions, which are closer to QCP, can be classified as 
LRMO systems. The magnetic structure of these compositions are presumably of an
antiferromagnetic type and not of a ferromagnetic type as 
indicated by the isothermal M behavior at 2 K and dc $\chi$ behavior 
reported in Ref. 1. 

As one decreases x to 0.4, there is a well-defined 
cusp in $\chi$$\prime$ at T$_o$ (Fig 1, top). 
Apparently there is a significant frequency
dependence of the values of $\chi$$\prime$ as well at a given T below and near the peak 
temperature. In addition, there is an upturn of $\chi$$\prime$$\prime$ below
3.5 K, which is most promiment for this composition (see Fig. 2). 
$\chi$$\prime$$\prime$
is also found to show marked frequency dependence as in the
case of $\chi$$\prime$; however,  for the sake of clarity, we have shown the
plot only at one frequency (1 Hz).   
These features reveal that this alloy undergoes SG freezing below T$_o$.
As the composition is varied towards Pd rich end, say for x= 0.2, 
the peaks  are broadened 
(Figs. 1 and 2) with an observable, but
a relatively weak, frequency dependence.  The features, 
particularly in $\chi$$\prime$$\prime$, are however relatively suppressed for x= 0.0.  
We would like to add that
ac $\chi$ measurements were carried on the single crystals (same crystal as in Ref. 3) and we are able to see 
spin-glass anomalies not only in $\chi$$\prime$ but also in $\chi$$\prime$$\prime$ (not shown here); in accordance with this, we have also noted the
separation of zero-field-cooled and field-cooled dc $\chi$ curves (H= 100 Oe) below about 2.3 K in this crystal. Hence, we believe
that these features are intrinsic to this compound and it therefore  appears that  the feature in $\chi$$\prime$$\prime$ in polycystals may be 
sometimes suppressed due to random orientations of the crystallites. 
It may be recalled that neutron diffraction pattern\cite{3} however revealed the features due to   antiferromagnetic 
ordering, but  not extending to the
entire crystal for x= 0.0;  magnetic cluster size has been found to be of the order
 of 100 $\AA$. Therefore, we believe that the  frequency dependence, particularly
for the Pd end member, arises from some degree of randomness of magnetic coupling among 
such clusters. In short, the Pd-rich alloys may be classified as cluster spin-glass systems.

To conclude, the results presented in this article suggest that, 
in the series, Ce$_2$Pd$_{1-x}$Co$_x$Si$_3$, the compositions which are 
far away from QCP, appear to behave like  (cluster) spin-glasses.  It is thus interesting to note that, as one
approaches QCP, the SG freezing in fact disappears in favor of LRMO.
It is of interest to perform neutron diffraction measurements to verify this observation considering its
implications.\cite{8}                    

The participation of Kartik K Iyer in magnetization measurements is gratefully 
acknowledged.


\begin{figure}


\caption{Real part of ac susceptibility of the alloys,   
Ce$_2$Pd$_{1-x}$Co$_x$Si$_3$, as a function of temperature
at various frequencies. For x= 0.6 and 0.7, the curves
at all the four frequencies overlap.  For the sake of clarity,
the data points have been omitted and a line  through
the data points is only shown for each curve.}

\end{figure}

\begin{figure}


\caption{Imaginery part of ac susceptibility of the alloys,   
Ce$_2$Pd$_{1-x}$Co$_x$Si$_3$, as a function of temperature
measured at a frequency of 1 Hz. For x= 0.6 and 0.7, the curves
overlap.  For the sake of clarity,
the data points have been omitted and a line  through
the data points is only shown for each curve.} 

\end{figure}


\begin{references}
\bibitem{6}A reinvestigation of poly- as well as single-crystals
of Ce$_2$PdSi$_3$, employed
in our earlier studies,\cite{1,3,4,5} by ac as well as dc $\chi$ studies
revealed that there is a weak shoulder around 4 K (see, for instance,
the curve for x= 0.0 in Fig. 1), as though there 
is a magnetic transition at this temperature in addition to the 
prominent one around 2.7 K.
\end{references}
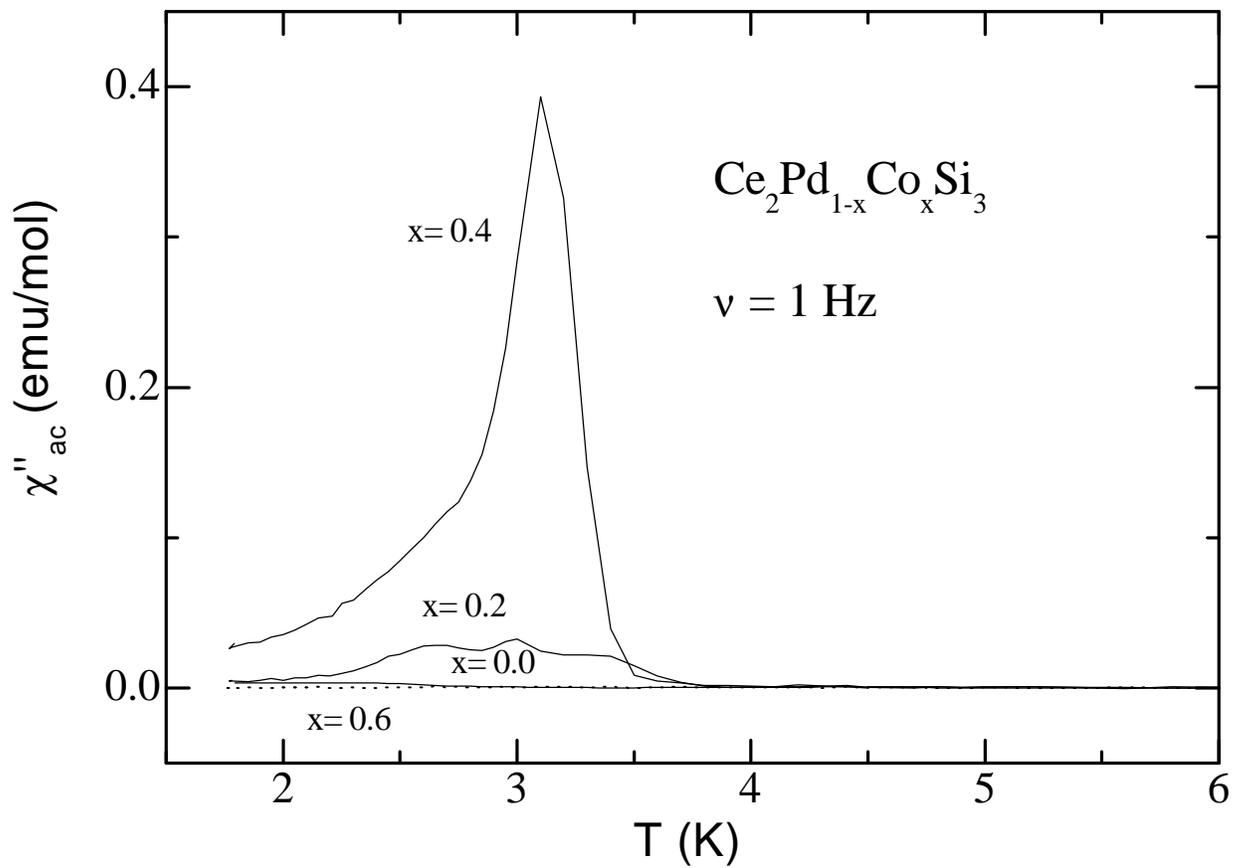
\end{document}